\documentclass[aps,prl,twocolumn,amsmath,amssymb,superscriptaddress]{revtex4-2}

\draft 

\usepackage[pdftex]{graphicx}
\usepackage{amsmath}
\usepackage{xr}

\begin{document}


\title{On Real-time Image Reconstruction with Neural Networks for MRI-guided Radiotherapy} 



\author{David E. J. Waddington}
\email[]{david.waddington@sydney.edu.au}
\affiliation{ACRF Image X Institute, Faculty of Medicine and Health, The University of Sydney, Sydney, Australia}
\affiliation{Department of Medical Physics, Ingham Institute for Applied Medical Research, Liverpool NSW, Australia}
\affiliation{A. A. Martinos Center for Biomedical Imaging, Massachusetts General Hospital, Charlestown, MA, USA}

\author{Nicholas Hindley}
\affiliation{ACRF Image X Institute, Faculty of Medicine and Health, The University of Sydney, Sydney, Australia}
\affiliation{A. A. Martinos Center for Biomedical Imaging, Massachusetts General Hospital, Charlestown, MA, USA}

\author{Neha Koonjoo}
\affiliation{A. A. Martinos Center for Biomedical Imaging, Massachusetts General Hospital, Charlestown, MA, USA}

\author{Christopher Chiu}
\affiliation{ACRF Image X Institute, Faculty of Medicine and Health, The University of Sydney, Sydney, Australia}

\author{Tess Reynolds}
\affiliation{ACRF Image X Institute, Faculty of Medicine and Health, The University of Sydney, Sydney, Australia}

\author{Paul Z. Y. Liu}
\affiliation{ACRF Image X Institute, Faculty of Medicine and Health, The University of Sydney, Sydney, Australia}
\affiliation{Department of Medical Physics, Ingham Institute for Applied Medical Research, Liverpool NSW, Australia}

\author{Bo Zhu}
\affiliation{A. A. Martinos Center for Biomedical Imaging, Massachusetts General Hospital, Charlestown, MA, USA}

\author{Danyal Bhutto}
\affiliation{A. A. Martinos Center for Biomedical Imaging, Massachusetts General Hospital, Charlestown, MA, USA}
\affiliation{Department of Biomedical Engineering, Boston University, Boston, MA, USA}

\author{Chiara Paganelli}
\affiliation{Dipartimento di Elettronica, Informazione e Bioingegneria, Politecnico di Milano, Milan, Italy}

\author{Paul J. Keall}
\affiliation{ACRF Image X Institute, Faculty of Medicine and Health, The University of Sydney, Sydney, Australia}
\affiliation{Department of Medical Physics, Ingham Institute for Applied Medical Research, Liverpool NSW, Australia}

\author{Matthew S. Rosen}
\affiliation{A. A. Martinos Center for Biomedical Imaging, Massachusetts General Hospital, Charlestown, MA, USA}
\affiliation{Department of Physics, Harvard University, Cambridge, MA, USA}
\affiliation{Harvard Medical School, 25 Shattuck St., Boston, MA, USA}


\date{\today}

\begin{abstract}

\textbf{Purpose:} MRI-guidance techniques that dynamically adapt radiation beams to follow tumor motion in real-time will lead to more accurate cancer treatments and reduced collateral healthy tissue damage. The gold-standard for reconstruction of undersampled MR data is compressed sensing (CS) which is computationally slow and limits the rate that images can be available for real-time adaptation.

\textbf{Methods:} We use automated transform by manifold approximation (AUTOMAP), a generalized framework that maps raw MR signal to the target image domain, to rapidly reconstruct images from undersampled radial k-space data. The AUTOMAP neural network was trained to reconstruct images from a golden-angle radial acquisition, a benchmark for motion-sensitive imaging, on lung cancer patient data and generic images from ImageNet. Model training was subsequently augmented with motion-encoded k-space data derived from videos in the YouTube-8M dataset to encourage motion robust reconstruction.

\textbf{Results:} AUTOMAP models fine-tuned on retrospectively acquired lung cancer patient data reconstructed radial k-space with equivalent accuracy to CS but with much shorter processing times. Validation of motion-trained models with a virtual dynamic lung tumor phantom showed that the generalized motion properties learned from YouTube lead to improved target tracking accuracy.

\textbf{Conclusion:} AUTOMAP can achieve real-time, accurate reconstruction of radial data. These findings imply that neural-network-based reconstruction is potentially superior to alternative approaches for real-time image guidance applications.

\end{abstract}
\maketitle 

\section{Introduction}
\label{intro}
Image-guided radiotherapy is a pillar of modern cancer treatment as it enables the noninvasive treatment of tumors with millimeter-scale accuracy while causing minimal damage to surrounding healthy tissue.\cite{Barton2014} At the cutting-edge of radiation oncology is a treatment machine known as an MRI-Linac, which combines the unrivaled image quality of magnetic resonance imaging (MRI) with linear accelerator (Linac) x-ray radiation therapy.\cite{Keall2022} Commercial MRI-Linacs are already achieving new standards of precision radiotherapy through image-guided adaptation to daily anatomical changes,\cite{VanSornsendeKoste2018} with cutting-edge developments including the implementation of gating techniques that dynamically shutter the radiation beam to account for patient motion.\cite{Stark2020} The next generation of MRI-Linac technology promises to track tumor motion with a moving radiation beam on the basis of real-time MRI.\cite{Keall2021} However, the accuracy of these targeting approaches, which are likely to improve patient outcomes and reduce side effects,\cite{Colvill2015} is limited by the low spatio-temporal resolution of MRI.\cite{Liu2020}

Fast MRI acquisitions based on acquiring raw k-space data with sparsely sampled golden-angle radial trajectories have shown much promise for tumor tracking during MRI-Linac treatments. These radial trajectories are unique in enabling reconstruction of high-spatial-resolution, motion-robust images\cite{Kumar2017} in parallel with high-temporal-resolution images from the same raw data.\cite{Bruijnen2019} Implementation of such imaging strategies would be advantageous in the radiotherapy context where low latency imaging is often reluctantly prioritized over resolution.\cite{Mickevicius2019} However, gold-standard techniques for analytic reconstruction of such undersampled MRI data are computationally slow, presenting a barrier to the real-time imaging required for dynamic treatment adaptation.\cite{Feng2016,Paulson2020} 

\begin{figure}
 \includegraphics{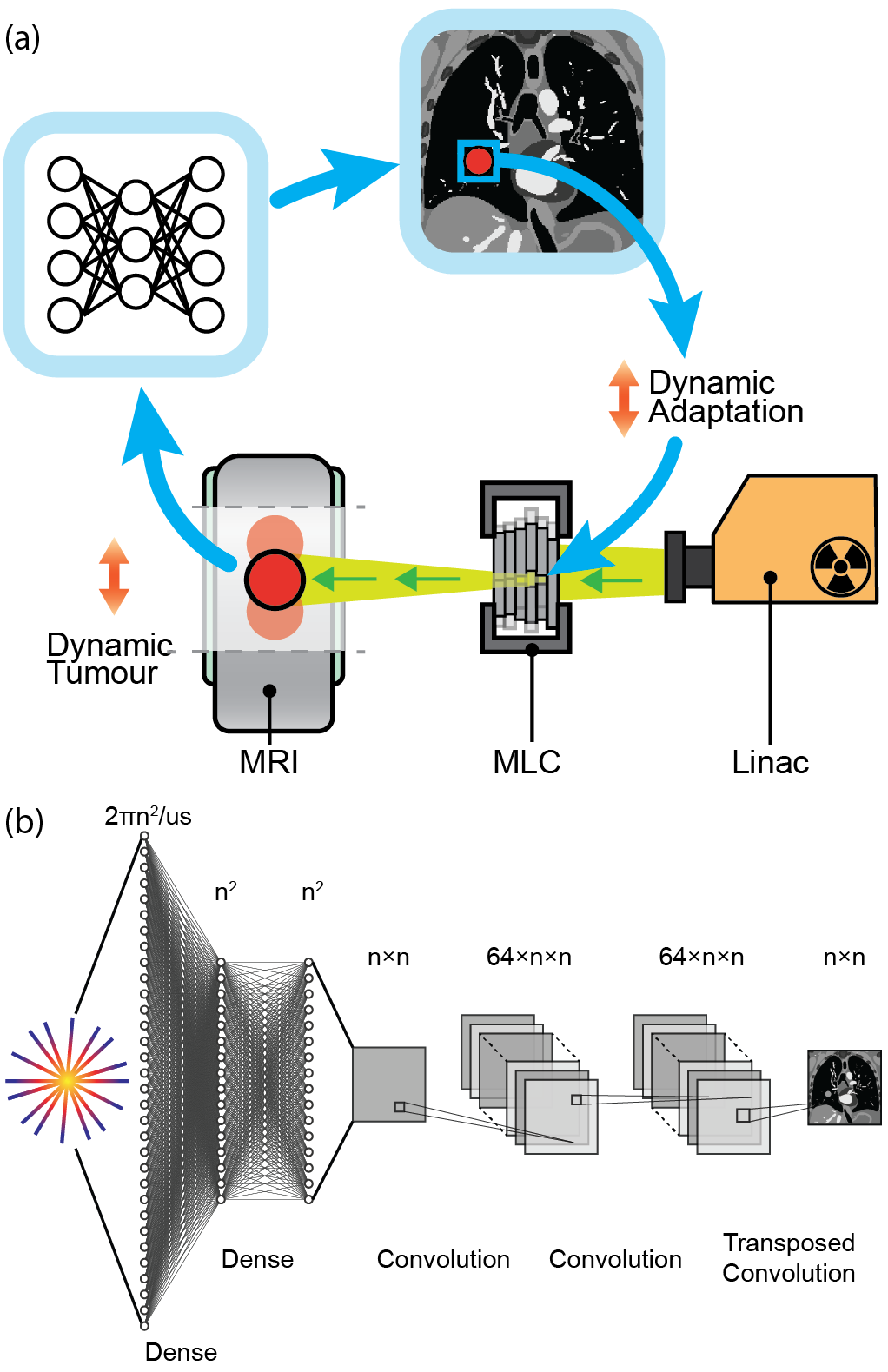}
\caption{\label{fig-vision} 
	\textbf{Deep neural networks as a fast, accurate reconstruction technique for tumor tracking applications.}
	\textbf{(a)} Workflow showing the potential role of AUTOMAP in a radiotherapy treatment with dynamic beam adaptation. Dynamic MRI scans are acquired on an MRI-Linac and reconstructed in real-time with AUTOMAP. A template-matching algorithm extracts the target position from images and dynamically adapts the X-ray beam via a multi-leaf collimator (MLC). 
	\textbf{(b)} The deep neural network architecture implemented to reconstruct an $n\times n$ image from radially-sampled MRI data with AUTOMAP. Radial k-space data is flattened into a 1D vector to create the input to a series of dense and convolutional layers that reconstruct an image.}
\end{figure}

Deep neural networks have fueled recent progress in computer vision, leading to new technologies across diverse fields such as autonomous vehicles,\cite{Feng2021} molecular analysis,\cite{Zhong2021} and medical imaging.\cite{Shad2021,Huynh2020} Common to many of these technologies is the requirement for pipelines that convert information acquired in an abstract sensor domain to an interpretable format on which real-world actions can be based. Recently, neural networks have enabled fast, accurate reconstruction of undersampled MRI data.\cite{Chandra2021,Wang2020} However, despite these prospects, the successful deployment of neural networks for real-time imaging applications on systems including MRI-Linacs (see Fig. \ref{fig-vision}a) still hinges on the availability of training data and utilization of a reconstruction framework suitable for more challenging, non-uniformly sampled image reconstruction.\cite{Park2020,Terpstra2020,Schlemper2019a,Li2019}. In particular, there is a dearth of training data for acquisitions corrupted by nonrigid motion, as a static ground truth does not exist.\cite{Johnson2019,Haskell2019,Lee2020}

One approach to performing real-time radial reconstruction is the use of automated transform by manifold approximation (AUTOMAP), a generalized neural-network reconstruction framework that learns the transformation from the raw MR signal to the target image domain from a training corpus built using the forward-encoding model.\cite{Zhu2018} Once trained, this machine-learning-based framework reconstructs images in a single forward pass.

Here, we train AUTOMAP to reconstruct undersampled golden-angle radial trajectory MR data. Using retrospectively acquired data from lung cancer patients, we compare the performance of AUTOMAP to conventional iterative methods for compressed sensing (CS) reconstruction, showing AUTOMAP gives similar reconstruction accuracy but with much faster processing times. Further, we leverage the YouTube-8M database to synthesize radial k-space data acquired in the presence of generic motion but with a known ground truth. We then show that our motion-trained AUTOMAP model leads to more accurate tumor targeting in a digital lung cancer phantom.\cite{Paganelli2017} These results will guide the development of neural network reconstruction techniques for low-latency, high accuracy reconstruction in real-time adaptive radiotherapy.

\section{Methods}
\label{methods}

\subsection{Model Architecture and Training Hyperparameters}

We implemented AUTOMAP using the architecture shown in Fig. \ref{fig-vision}b with Keras (2.4.3) operating on a TensorFlow backend (2.5.0).\cite{Zhu2018} The AUTOMAP architecture was created with 5 trainable layers. The first two layers are dense with hyperbolic tangent activations and map flattened input data through hidden $n^2 \times 1$ layers. Data is reshaped to an $n \times n$ matrix. Data then passes through two convolutional layers with 64 filters, kernel size $5 \times 5$ and rectified linear activation functions before a transposed convolution with 1 filter and a $7 \times 7$ kernel produces the final $n \times n$ image. Models were trained to reconstruct images with $n = 128$, with a different model for each acceleration factor.

Training utilized the Adaptive Moment Estimation (Adam) optimizer with a learning rate of 10$^{-5}$, a batch size of 20 and a root mean square error (RMSE) loss function. Model weights corresponding to the minimum validation cost achieved in 300 epochs of training, with a patience of 15, were saved for reconstruction. For reconstruction of lung cancer patient images, models trained on ImageNet were fine-tuned for up to 100 epochs with a learning rate of 10$^{-6}$. All computation utilized a NVIDIA RTX 8000 or NVIDIA A6000 graphical processing unit (GPU) on an Ubuntu 18.04 workstation with 128~GB RAM and a 10-core 3.5~GHz Intel central processing unit (CPU).

\subsection{Image Data Preprocessing}

We begin by training AUTOMAP to perform image reconstruction under the assumption that anatomy is static. For initial model training, datasets of 20,000 training images and 1,000 validation images depicting generic objects were sourced from ImageNet and augmented 4 times via a series of flips and rotations.\cite{Deng2009} ImageNet data were converted to normalized, grayscale images at 256$\times$256 resolution and augmented further via addition of random synthetic phase maps. We highlight that data augmentation with these phase maps, which are fully described in Ref \citenum{Zhu2018}, is essential to preventing AUTOMAP from overfitting during training.

Using the MATLAB toolbox for realistic analytical phantoms described in Ref. \citenum{Guerquin-Kern2012}, the ImageNet derived datasets were encoded with the golden-angle radial trajectories to generate single coil k-space datasets via a Nonuniform Fast Fourier Transform (NUFFT) operation. Golden-angle trajectories for 128$\times$128 resolution reconstruction were undersampled in the phase-encode direction by reduction factors (\textit{R}) of 1-16 and oversampled 2-times in the frequency-encode direction. The $256 \times 256$ grayscale images were downsampled via bilinear interpolation to $128 \times 128$ for use as ground truth target images in model training.

For model fine-tuning and evaluation, MRI scans from a 13 patient lung cancer dataset were split in the ratio 9/2/2 (training/validation/testing). This lung cancer dataset is fully described in Refs \citenum{Lee2017} and \citenum{Lee2018a}. Radial, single-coil k-space data for these images were retrospectively acquired from images saved in the Digital Imaging and Communications in Medicine (DICOM) format with the procedure described above. Slices from T1-weighted, T2-weighted and cine-MRI scans were analyzed individually, yielding several hundred images per patient. To simulate real-world sensor noise, additional k-space datasets with 25~dB of additive white Gaussian noise (AWGN) was added created with the Signal Processing Toolbox in MATLAB.

\subsection{Motion Data Preprocessing}

The next part of work this aims to account for intra-acquisition motion in the neural network reconstruction by incorporating generic motion into the training corpus. Here, we aim to make the motion-correction generalize by using videos from YouTube as a source of motion-encoded training data.

For motion training, 7,856 image sequences were extracted from 767 videos in the `wildlife' class of the YouTube-8M database and split randomly into training/validation sets at the ratio 0.85/0.15.\cite{Abu-El-Haija2016} The extraction process ensured that the video sequences contained continuous, smooth motion by excluding cases where the structural similarity (SSIM) between any adjacent frames was lower than 0.94. Sequences where the SSIM between first and last frames was lower than 0.5 were also excluded. These thresholds were chosen as typical of the SSIM values observed across the cine-MRI lung data described above.

\begin{figure*}
	\includegraphics{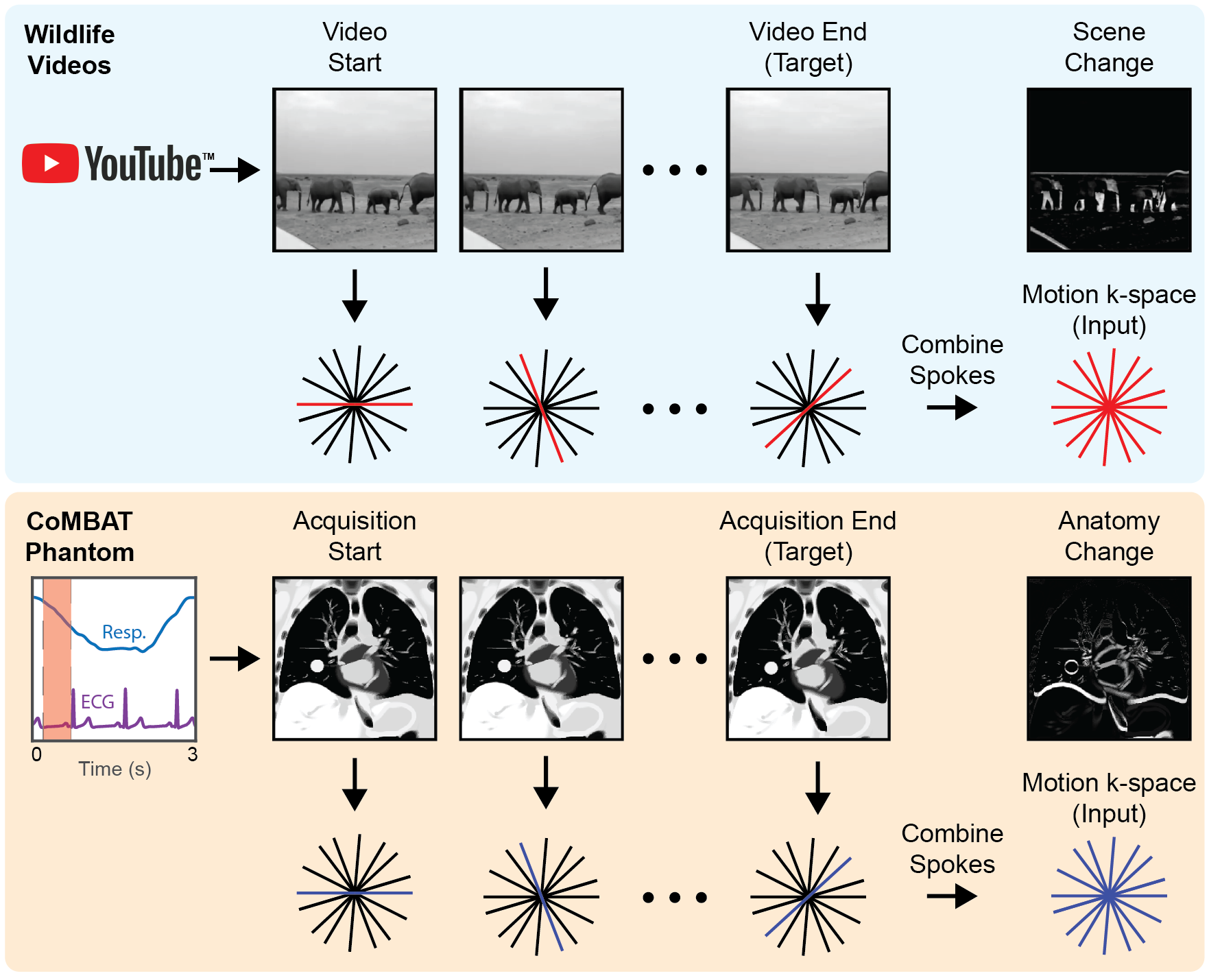}
	\caption{\label{fig-motion-encoding} 
		\textbf{Simulating patient motion during radial acquisitions.}
		The CoMBAT phantom inputs respiratory and ECG traces to simulate patient anatomy during cardiothoracic motion. MR slices are simulated at each timepoint during the acquisition (red shading) and encoded to a golden-angle radial trajectory. A `motion-encoded' k-space is derived by taking individual spokes from the `static' k-space at individual timepoints during the acquisition. The anatomy change between start and end timepoints is shown as a difference image.
	}
\end{figure*}

Motion-encoded radial k-space data was created from image sequences by combining spokes of readout data from sequential `static' 4-times undersampled k-space data generated with the NUFFT procedure described above (see Fig. \ref{fig-motion-encoding}). AWGN at 25~dB was added to raw k-space data to simulate sensor noise. For ground truth data, we selected the image corresponding to the last frame in each k-space acquisition because knowledge of the most recent anatomical state is desired for real-time beam adaptation on an MRI-Linac.

Motion sequence data was augmented by a factor of 8 via a series of flips, rotations and time-reversal processes. Data were further augmented by the addition of random phase maps. For final model training, this YouTube-8M dataset was combined with the ImageNet dataset described above. A separate motion test set of 1000 motion sequences and input data were created from an independent 218 videos in the YouTube-8M dataset using the SSIM criteria described above.

As a testing tool, a time-series of 2D lung cancer images were generated using the digital CT/MRI breathing XCAT (CoMBAT) phantom for a balanced steady-state free precession (bSSFP) sequence with TR/TE = 10/5 ms.\cite{Paganelli2017,Segars2010,Reynolds2019a} Image sequences were transformed to $4\times$ undersampled k-space data using the process shown in Figure \ref{fig-motion-encoding}.

\subsection{Image Reconstruction}

Neural network image reconstruction was performed by running inference on flattened radial input data with the corresponding, trained AUTOMAP model. The AUTOMAP reconstruction time was measured as wall time taken to perform this inference step in Keras on an unburdened workstation as measured over 20 repeats.

Conventional compressed sensing (CS) and non-uniform fast Fourier transform (NUFFT) image reconstruction techniques were performed using the Berkeley Advanced Reconstruction Toolbox (BART) toolbox.\cite{BART-toolbox,Tamir2016} The NUFFT reconstruction interpolates k-space data onto a Cartesian grid and then performs a fast Fourier transform.\cite{Fessler2003} The CS implementation utilizes a NUFFT with an iterative algorithm to find the $l1$-regularized solution to:
\begin{equation}
			\min_{x} \left\{ \| \mathbf{A} \mathbf{x} - \mathbf{y} \|_2 + \lambda \| \psi \mathbf{x} \|_1 \right\},
\end{equation}
where $\mathbf{y}$ is the acquired k-space, $\mathbf{A}$ is the (under)sampling operator over the reconstructed image $ \mathbf{x}$, $\lambda$ is a regularization parameter and $\psi$ is the wavelet operator. A grid search was used to optimize $\lambda$ in the range $10^{-7} - 10^{-1}$ for a minimum normalized root mean square error (NRMSE) with 30 iterations. Reconstruction times for CS are the self-reported time for reconstruction as measured by the BART toolbox over 20 repeats.

The NRMSE was used as the primary metric to evaluate reconstruction quality and is calculated as the RMSE between reconstructed image and ground truth divided by the intensity range of the ground truth image. Structural similarity (SSIM) and the Peak signal-to-noise ratio (PSNR) are also considered as additional quantitative metrics that may indicate the clinical utility of the images.\cite{Mason2020} In Figure \ref{fig-metrics} the values reported for SSIM/NRMSE are the mean and standard deviation across 400 image slices in the 2 patient subset of lung cancer dataset that was unseen by AUTOMAP.

\subsection{Template Matching}
To simulate target tracking, regions of interest encompassing the tumor and diaphragm were defined in the first ground truth image of the digital phantom. Using CS and trained neural networks, stacks of 240 images were reconstructed from data retrospectively acquired with cardiothoracic motion. A template matching algorithm based on OpenCV software then calculated the closest matching target location from these reconstructed images with a normalized cross-correlation algorithm at half-pixel resolution (1~mm).\cite{Liu2020}

\section{Results}
\label{results}

\subsection{Model Training}

Models trained for static reconstruction on ImageNet derived data converged smoothly for all undersampling factors tested, with time per epoch being in the range 5 min. (\textit{R}~=~16) to 15 min. (\textit{R}~=~1). A plateau in validation cost was found within 300 epochs of training for networks with \textit{R}~$>$~4. Models pre-trained on ImageNET data were then fine-tuned on a lung cancer images. For example, with \textit{R}~=~4, the fine-tuned AUTOMAP model had a validation cost 9.7 times lower than a model trained from scratch on the lung cancer training data, emphasizing the value of transfer learning from generic pre-trained models when only smaller datasets are available for a given anatomy. Examples of training/validation cost training dynamics and further analysis are included in Supplementary Figure \ref{suppfig-training}.

\subsection{Neural Network Reconstruction Performance}
\label{sec_nn_perf}

\begin{figure*}
	\includegraphics{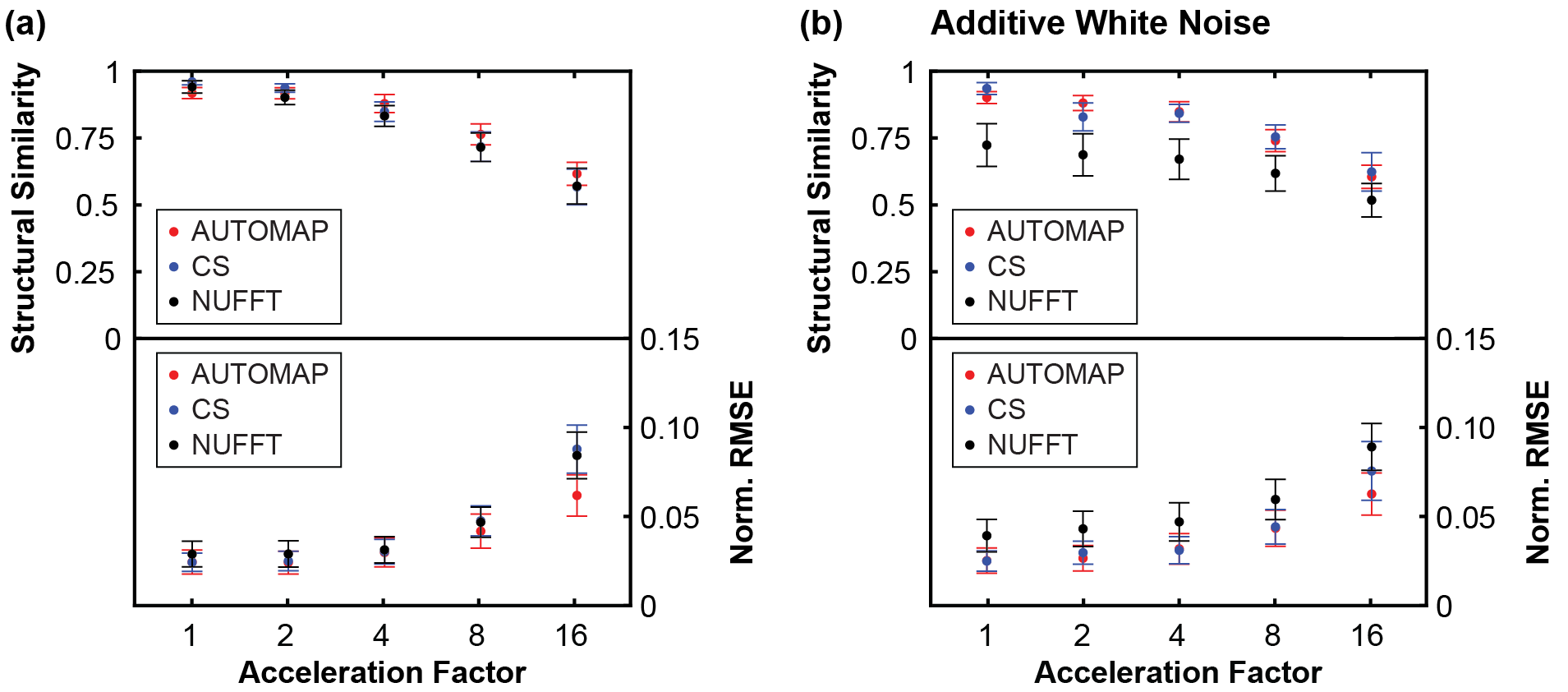}
	\caption{\label{fig-metrics} 
		\textbf{Quality of AUTOMAP reconstruction in comparison to conventional image reconstruction techniques for radial sampling.} Reconstruction quality as measured via structural similarity and normalized root mean square error metrics (Norm. RMSE) for different undersampling factors. Results are shown for data reconstructed with AUTOMAP (red), compressed sensing (CS, blue) and non-uniform fast Fourier transform (NUFFT, black) techniques. Resulting image quality was assessed for clean radial data retrospectively encoded from images (shown in \textbf{a}) and for the same data with 25~dB of additive white Gaussian noise (shown in \textbf{b}).
	}
\end{figure*}

\begin{figure*}
	\includegraphics{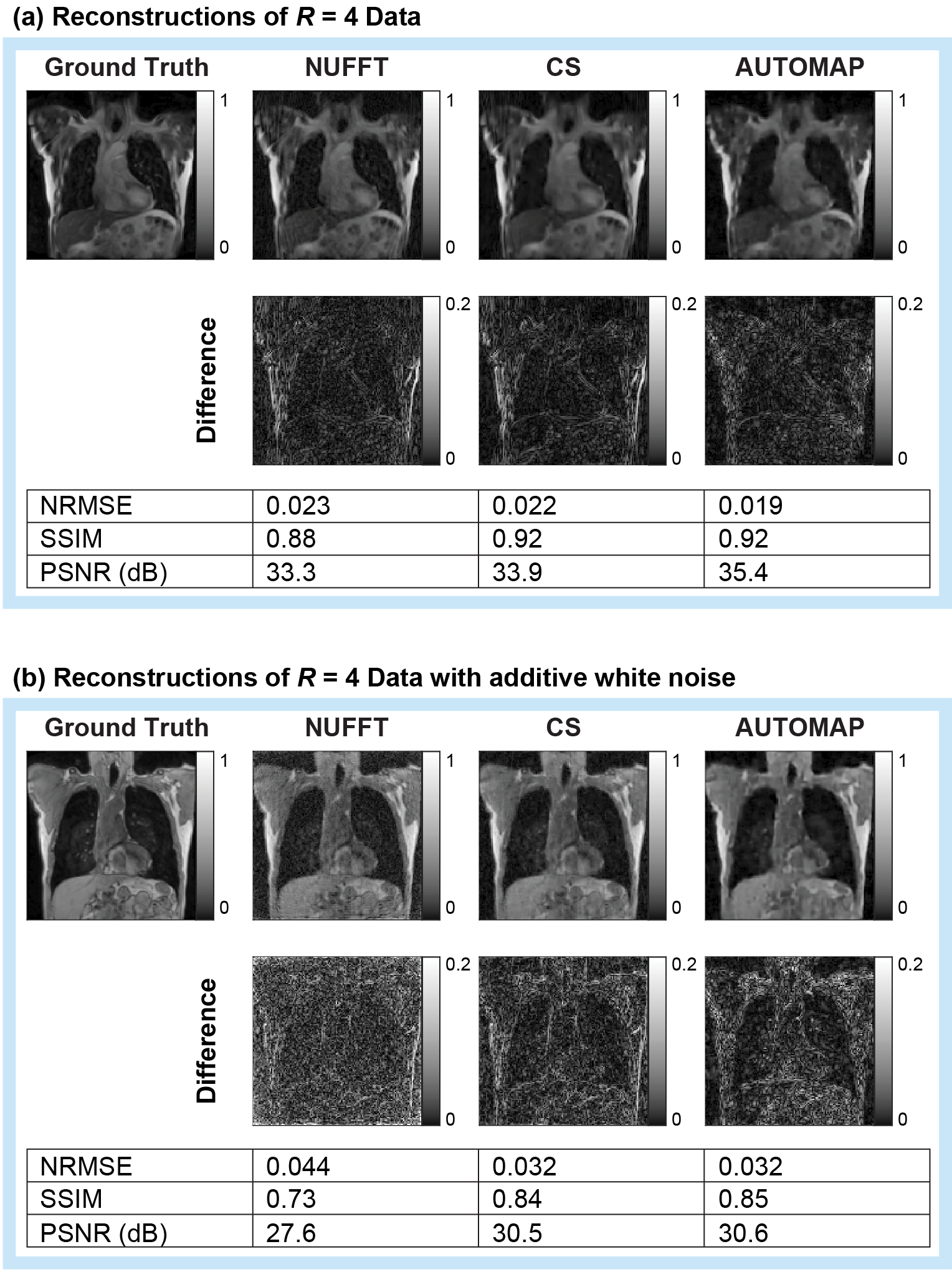}
	\caption{\label{fig-static-recons}
		\textbf{Visual comparison of AUTOMAP reconstruction performance to conventional techniques.}
		Reference images from a lung cancer patient dataset were encoded to a 4$\times$ undersampled golden-angle radial trajectory and reconstructed with AUTOMAP, compressed sensing (CS) and non-uniform fast Fourier transform (NUFFT)  techniques. 
		Resulting image quality was assessed for clean radial data retrospectively encoded from images (shown in \textbf{a}) and for the same data with 25~dB of additive white Gaussian noise (shown in \textbf{b}).
		Error maps show the difference between reconstructed and ground truth images. Normalized root-mean-square error (NRMSE), structural similarity (SSIM) and peak signal-to-noise ratio (PSNR) image quality metrics are shown. 
	}
\end{figure*}

With models trained for the neural network reconstruction task, we now present key results in Figures \ref{fig-metrics} and \ref{fig-static-recons} that compare the reconstruction error of AUTOMAP to established methods.

In Fig. \ref{fig-metrics}, we show SSIM and NRMSE metrics for images reconstructed from golden-angle radial k-space data derived from the lung cancer imaging  test set at a range of acceleration factors ($R$). AUTOMAP performs strongest with very sparsely sampled data, giving an NRMSE 30\% lower than CS reconstruction for $R~=~16$ data as shown in Fig. \ref{fig-metrics}a. We observe that NUFFT reconstructs the `clean' data effectively, having an NRMSE only 4\% higher than CS for $R~=~4$ (Fig. \ref{fig-metrics}a). However, when white noise, representing the thermal noise present in a standard MRI  experiment,\cite{Redpath1998} is added to the input k-space data, we observe considerable deterioration in the accuracy of NUFFT reconstructions performed with the NRMSE being 52\% higher than for CS at $R~=~4$ (Fig. \ref{fig-metrics}b). For fully sampled data ($R~=~1$), CS and AUTOMAP give an NRMSE within 1\% even after white noise is added.

Aware that NRMSE and SSIM metrics do not fully characterize artifacts that may be present in reconstructed images, we present typical reconstructions for \textit{R}~=~4 shown in Fig. \ref{fig-static-recons}. While, the overall quality trend is similar to that summarized in Fig. \ref{fig-metrics}, we also observe that the AUTOMAP difference image is distinguished by Moir\'e-like features while edges predominate in the CS difference image. Increasing the regularization penalty was observed to remove structure from the CS difference images at the expense of a higher NRMSE.

Having shown that AUTOMAP reconstructs input data with equivalent or superior fidelity to CS, we now evaluate the relative computational performance of the reconstruction methods. We found that our implementation of AUTOMAP was 16-49 times faster than CS reconstruction, reducing reconstruction time by approximately 200~ms (see Table \ref{recontimes}). For context, the end-to-end imaging and targeting latency should be less than $\sim$500~ms for real-time tumor tracking in MR-guided radiotherapy.\cite{Keall2021} The speed of NUFFT reconstruction was not quantified, as it is accepted that this less robust reconstruction technique can be completed within several milliseconds, making it sufficiently fast for real-time applications.\cite{Lin2018} We found that CS reconstruction was fastest running on the CPU without GPU acceleration, presumably due to the relatively small matrix sizes used, and hence, report the CPU-based times here.

\begin{table*}[]
	\begin{tabular}{ll|lllll}
		\hline
		\textbf{Acceleration Factor (\textit{R})}                          &                    & 1            & 2            & 4            & 8           & 16          \\ \hline \hline
		\multicolumn{1}{l|}{}                         & AUTOMAP 1-Slice    & 16.0$\pm$0.3 & 10.4$\pm$0.4 & 7.2$\pm$0.6  & 5.3$\pm$0.7 & 4.7$\pm$0.9 \\
		\multicolumn{1}{l|}{\textbf{Reconstruction Time (ms)}} & AUTOMAP 10-Slices  & 19.1$\pm$0.4 & 13.5$\pm$0.4 & 10.0$\pm$0.4 & 8.2$\pm$0.7 & 7.1$\pm$0.8 \\
		\multicolumn{1}{l|}{}                         & Compressed Sensing 1-slice & 253$\pm$23   & 240$\pm$17   & 235$\pm$8    & 236$\pm$10  & 231$\pm$21  \\ \hline
		\textbf{Number of AUTOMAP Parameters}                  &                    & 1.96B        & 1.12B        & 696M         & 487M        & 378M        \\ \hline
		\textbf{CS Regularization Parameter ($\lambda$)})                  &                    & 0.005        & 0.005        & 0.005         & 0.01       & 0.05        \\ \hline
	\end{tabular}
	\caption{\label{recontimes} Computational load of image reconstruction techniques. The time to perform image reconstruction with AUTOMAP and compressed sensing techniques for different undersampling factors is shown in addition to the number of trainable parameters in AUTOMAP. Compressed sensing reconstructions utilize one optimized hyperparameter value for each undersampling factor.}
\end{table*}

The number of trainable parameters required to implement AUTOMAP were significant (Table \ref{recontimes}), reaching up to 2B for the fully-sampled network. A single $l1$-penalty hyperparameter ($\lambda$) was optimized for CS recon.

\subsection{Motion-compensated Reconstruction}

Here, we test the potential of the AUTOMAP reconstruction technique to correct for anatomical motion encountered during the acquisition process, which is of concern for MRI-guided RT of thoracic sites in particular. In this section we consider two AUTOMAP models trained to reconstruct data acquired with \textit{R}~=~4. The first AUTOMAP model is the same as the \textit{R}~=~4 model trained for reconstruction as described above but without fine-tuning. The second AUTOMAP model was trained to compensate for intra-acquisition motion through the incorporation of generic motion data from YouTube-8M into the training dataset.

\begin{figure*}
	\includegraphics{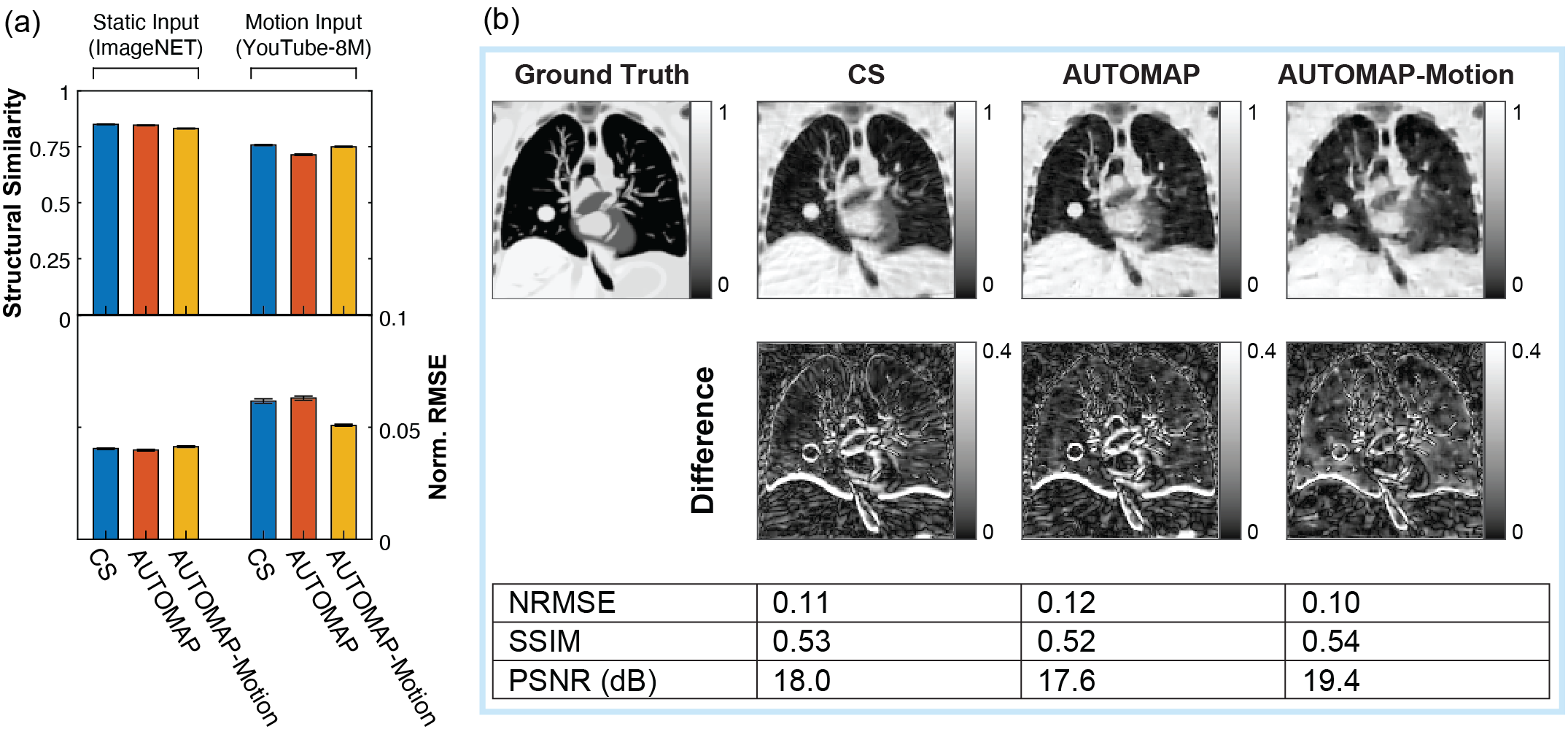} 
	\caption{\label{fig-motion-recon} 
	\textbf{Reconstructing motion-corrupted data.}
	\textbf{(a)} Reconstruction quality as measured via structural similarity and normalized root mean square error metrics (Norm. RMSE) for static images derived from the ImageNET database and for motion-encoded inputs derived from the YouTube 8M database. Results are shown for 4$\times$ undersampled data reconstructed with compressed sensing (blue), an AUTOMAP model trained on static data (red) and an AUTOMAP model trained on motion-encoded data. Bars and lines are the mean and standard error of the mean calculated across 1,000 test inputs.
	\textbf{(b)} Images reconstructed from k-space data simulated with the CoMBAT phantom for a patient under routine cardiothoracic motion. Results for data reconstructed with compressed sensing, an AUTOMAP model trained on static data and an AUTOMAP model trained on motion-encoded data are shown.
	Normalized root-mean-square error (NRMSE), structural similarity (SSIM) and peak signal-to-noise ratio (PSNR) image quality metrics are evaluated against the last frame of the image sequence for motion-encoded data.
	Additive white Gaussian noise at 25~dB was added to all inputs used to derive this figure.
}
\end{figure*}

We begin analyzing these models by comparing the quality of CS reconstruction to the AUTOMAP model reconstructions for static and motion inputs in Fig. \ref{fig-motion-recon}a. The quality of all reconstructions is higher for static input data than motion input data as would be expected, with the NRMSE for all static reconstructions being within 2.5\%. The best performing technique on motion input data was the motion-trained AUTOMAP model, which had a mean NRMSE 21\% lower than the CS reconstructions. Additionally, the performance of AUTOMAP was comparable on static input data whether or not the model was trained with the inclusion of motion data, indicating that motion training leverages previously underutilized capacity in the over-parameterized model architecture. Conversely, we note that AUTOMAP trained on the YouTube-8M dataset alone had a minimum validation loss 68\% higher than when static data was included in the training set, likely due to the difficulty of fitting to the underlying manifold with variability in the motion-encoded data.

Having evaluated the performance of these reconstruction techniques on generic motion sequences, we now turn to analyze the performance of these models in reconstructing motion-corrupted k-space data simulated from the CoMBAT phantom for a lung cancer patient undergoing realistic cardiothoracic motion.\cite{Garcia-Gonzalez2013} The reconstruction results for CoMBAT data encoded as per the process described in Fig. \ref{fig-motion-encoding} are shown in Figure \ref{fig-motion-recon}b. The quantitative metrics of reconstruction performance are consistent with the results in Figure \ref{fig-motion-recon}a, with the motion-trained AUTOMAP model outperforming the standard model and CS. Inspecting difference images, we see that the output of the motion-trained AUTOMAP model has the least discrepancy with the ground truth (anatomy at acquisition end) around the diaphragm and tumor.

\begin{figure*}
	\includegraphics{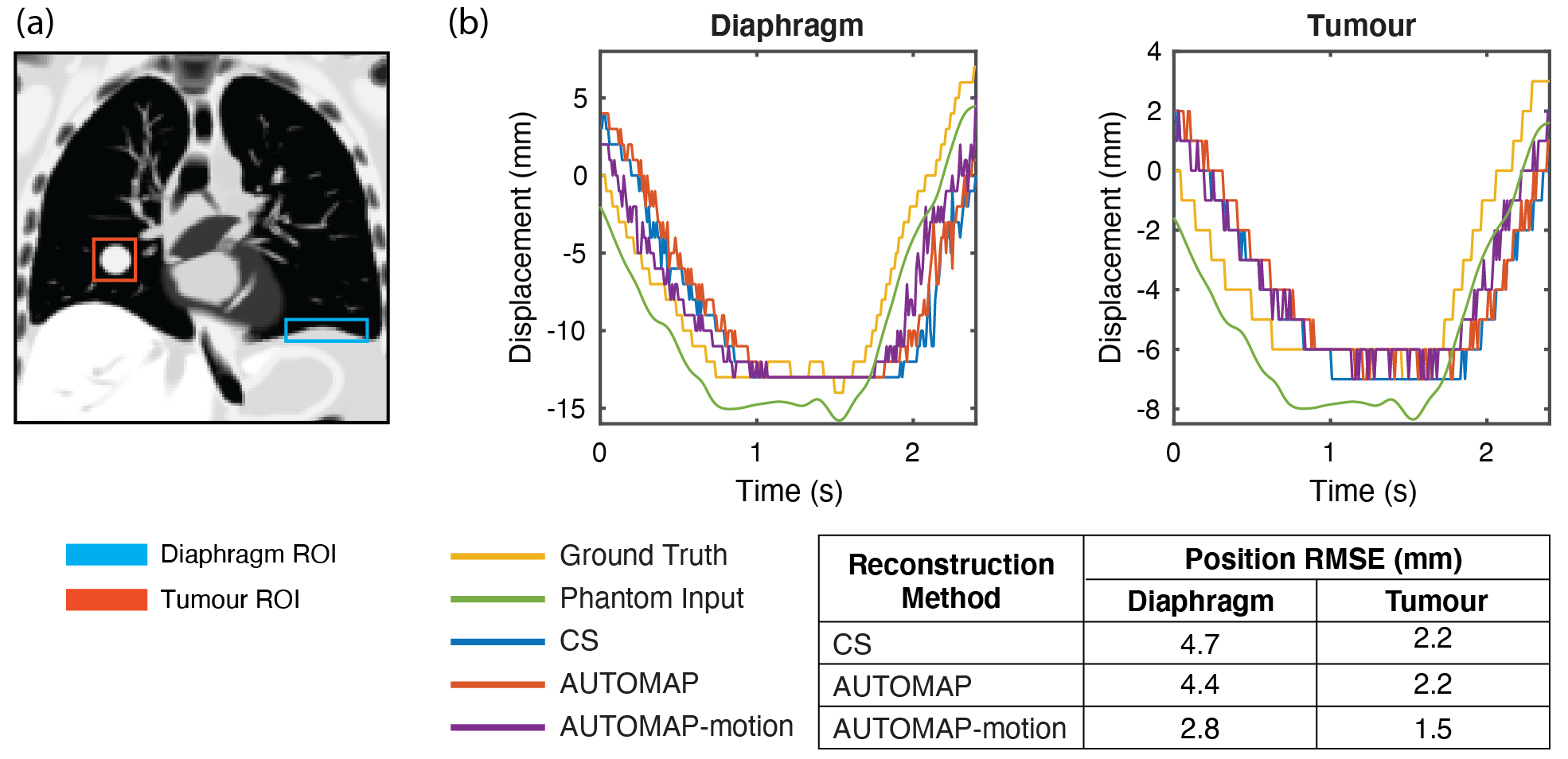} 
	\caption{\label{fig-tracking}
		\textbf{Target tracking accuracy.}
		\textbf{(a)} Regions of interest (ROIs) for the diaphragm (blue) and tumor (red) are defined in a ground truth image.
		\textbf{(b)} Displacement of ROIs defined in \textbf{a} as predicted by a template matching algorithm for the ground truth image sequence (yellow) and image sequences reconstructed using compressed sensing (CS, blue), a conventional AUTOMAP model (red) and an AUTOMAP-motion model (purple). Steps in displacement reflect the underlying image resolution. Root mean square error values are calculated for the difference between target position in reconstructed image sequences and the ground truth image sequence. The motion trace input to the virtual phantom is shown (green) with a vertical offset for visibility.
	}
\end{figure*}

To simulate the impact of motion-training on an MRI-Linac tracking experiment, we performed template matching of tumor and diaphragm ROIs in our digital phantom, as shown in Figure \ref{fig-tracking}. Analyzing the difference between template match locations in ground truth images and reconstructed images through one respiratory cycle, we find that the AUTOMAP-motion model gives an RMSE value 1.9~mm smaller for the diaphragm and 0.7~mm smaller for the tumor than CS.

\section{Discussion}
\label{discussion}

Our results leverage advances in machine learning to implement fast image reconstruction of undersampled radial data from lung cancer patients with comparable accuracy to slower, gold-standard iterative reconstruction techniques. While our proof-of-principle study has focused on the application of AUTOMAP to real-time targeting of radiotherapy in the lung, we believe our results are extensible to high-motion sites such as the liver and prostate, where tumor movement would be optimally managed by real-time adaptive radiotherapy.\cite{Bertholet2019,Booth2021} We note that due to the relatively high latency of MR acquisition and reconstruction, compared to X-ray-based modalities, faster image reconstruction techniques are desired for real-time beam gating and MLC tracking on MRI-Linacs, especially for non-Cartesian acquisition trajectories.\cite{Fast2019,Liu2020} Integrating neural networks with fast data streaming tools,\cite{Borman2019} tracking algorithms\cite{Toftegaard2018} and time-resolved 3D anatomical imaging\cite{Bauman2016, Rabe2021} will be crucial for use with MRI-Linac beam adaptation technologies. 

The ease with which AUTOMAP generalizes to different reconstruction tasks, like the golden-angle radial sampling used in this work, is a direct consequence of the sequential dense layers in the model architecture. While these dense layers enable data-driven learning of the manifold between k-space and the image domain, they also have significant memory requirements that make the translation to 3D reconstruction and tracking more challenging.\cite{Paganelli2019,Koonjoo2021} To perform reconstructions above the relatively low resolutions used for motion tracking on an MRI-Linac will require lighter-weight reconstruction networks.\cite{Kurz2020,Chandra2021} One light-weight implementation of AUTOMAP is decomposed-AUTOMAP (dAUTOMAP) which replaces dense layers with orthogonal `domain transform' layers.\cite{Schlemper2019} While dAUTOMAP performs strongly for Cartesian trajectories, it assumes that data is acquired in orthogonal directions, making it unsuitable for reconstruction of nonuniform data. Hence, dAUTOMAP is outperformed by NUFFT reconstruction for radial reconstructions.\cite{Terpstra2020} We note that light-weight architectures that take view angle into account are an active research area.\cite{Li2019,Yoo2021}

Without a detailed understanding of the AUTOMAP learning process, it may also seem surprising that the neural network trains faster and performs better relative to CS at high acceleration factors (see Fig. \ref{fig-metrics}). However, this reflects the underlying mathematics of AUTOMAP as a tool for manifold approximation, where image reconstruction is achieved via a mapping between sparse representations in signal and image space. Hence, highly undersampled data, encourages the network to learn robust, low-dimensional, latent representation of data that can then be used for manifold approximation.\cite{Vincent2008}

Our experiments with motion-encoded k-space demonstrated that as a highly over-parameterized model, AUTOMAP has significant capacity to learn additional features. Here, we showed that AUTOMAP could learn generic properties of motion from YouTube videos, leading to lower NRMSE and higher tracking accuracy in reconstructed images of an \textit{in silico} lung cancer phantom. These tracking results do not account for reconstruction latency, which would be expected to further reduce tracking accuracy with CS. Additionally, there is significant scope to improve on tracking performance through utilization of more in-domain training data via synthetically generated motion applied to lung images. Such an approach could be adapted to assist with tracking of targets with out-of-plane motion, which represents a persistent challenge in MRI-guided radiotherapy.\cite{VanDeLindt2021} Temporal k-space filtering,\cite{Bruijnen2019} optical flow techniques\cite{Terpstra2020} and neural networks tailored to radial reconstruction could also improve network performance.\cite{Wang2018, Liu2019a,Yoo2021}

Our results show that AUTOMAP performs well with the addition of white noise, which represents the fundamental thermal limitations to SNR in an MRI scan. However, another valuable strength of data-driven MRI reconstruction models such as AUTOMAP is that they implicitly learn to suppress common MRI artifacts, such as spike noise caused by RF leakage, as these inputs typically fall outside the training domain.\cite{Koonjoo2021}

Future work will be required to test the robustness of AUTOMAP to common experimental imperfections encountered in MRI, such as $B_0$ and $B_1$ artifacts. While the over-parameterization of AUTOMAP means that it can be intentionally trained to correct for such artifacts, the incorporation of such specific examples into the training corpus will increase the risk of overfitting. Despite the risk of overfitting, it is likely that performance of AUTOMAP will benefit from some fine-tuning of pre-trained networks to the particular MRI system being used, especially when extending the approach to more challenging multi-coil reconstructions.\cite{Knoll2020} The incorporation of new adversarial approaches into the training corpus will make neural network reconstructions more robust by identifying nonphysical input perturbations that can negatively impact reconstruction performance.\cite{Yu2019,Antun2020}

In conclusion, we have used AUTOMAP to accurately and rapidly reconstruct retrospectively acquired lung cancer images from radial data. We have also shown that AUTOMAP can adapt to generalized properties of motion learned from generic YouTube videos for real-time tracking applications. These results will inform the future development of dynamic adaptation technologies for MRI-Linacs, enabling new standards of personalized radiotherapy.

\section*{Acknowledgements}
D Waddington is supported by a Cancer Institute NSW Early Career Fellowship 2019/ECF1015. This work has been funded by the Australian National Health and Medical Research Council Program Grant APP1132471. The authors thank NVIDIA for providing the RTX 8000 GPU used in this work.

\section*{Data Availability}

The primary datasets used for training in this study are publicly available. Code for preprocessing data and model training are available at https://github.com/MattRosenLab.


%

\end{document}